\documentclass[11pt]{article}
\usepackage[margin=1in]{geometry}
\usepackage{times}
\usepackage{graphicx}
\usepackage{booktabs}
\usepackage{hyperref}
\usepackage{amsmath}
\usepackage{enumitem}
\usepackage{seqsplit}
\usepackage{authblk}
\hypersetup{colorlinks=true, linkcolor=blue, citecolor=blue, urlcolor=blue}

\title{\textbf{Characterizing the eBPF-Based Data Plane for Multi-Cluster Kubernetes: A Systematic Evaluation of Cilium Cluster Mesh and KVStoreMesh}}
\author[1]{Simhadri Podala Narasimha}
\affil[1]{Independent Researcher, Littleton, Colorado, USA \\ \texttt{podalas@gmail.com}}
\date{}

\begin{document}
\maketitle

\begin{abstract}
Kubernetes deployments increasingly span multiple clusters for reasons of scale, fault isolation, regulatory boundaries, and heterogeneous hardware placement, including GPU-dense clusters dedicated to AI and HPC workloads. Cilium Cluster Mesh, built on extended Berkeley Packet Filter (eBPF) technology, has emerged as a widely deployed mechanism for connecting such clusters into a single logical network without a dedicated multi-cluster gateway. Despite its adoption, the academic literature contains no systematic, reproducible evaluation of the eBPF-based multi-cluster data plane itself: existing work either benchmarks eBPF against iptables within a single cluster, or applies eBPF to cross-cluster monitoring and anomaly detection rather than to the forwarding and identity-propagation path. The most detailed scale data available -- a report of a clustermesh-apiserver deployment failing under load at approximately 45{,}000 nodes across 256 clusters -- comes from an industry engineering blog rather than a peer-reviewed source. This paper proposes a methodology to close that gap. We define five research questions covering cross-cluster forwarding latency and throughput, identity and endpoint propagation delay, control-plane scalability limits, behavior under adversarial churn, and sensitivity of GPU/RDMA traffic to cross-cluster forwarding, and we describe a testbed and measurement design -- including eBPF-level instrumentation via Hubble and custom probes -- for answering them. Because a live multi-cluster testbed was not available for this submission, we additionally report results from a calibrated analytical simulation of the two control-plane architectures, fit to the single real production data point available in the literature; these results are explicitly presented as falsifiable model-derived predictions rather than empirical measurements, and are intended to be confirmed, revised, or falsified once the proposed testbed is built. This paper presents motivation, related work, research questions, a full experimental design, and this simulation-based first look; it is intended as a template and preliminary result set for a measurement study to be executed in full and reported empirically in a follow-on submission.
\end{abstract}


\section{Introduction}
Organizations operating Kubernetes at scale rarely confine themselves to a single cluster. Regulatory boundaries, blast-radius containment, multi-region availability, and the physical placement of specialized hardware such as GPUs all push infrastructure teams toward federations of clusters that must nonetheless behave, from an application's perspective, like a single connected platform. Cilium, an eBPF-based Container Network Interface (CNI), addresses this need through a feature called Cluster Mesh, which synchronizes service, endpoint, and identity state across cluster boundaries and enables pod-to-pod connectivity, global service discovery, and consistent network policy enforcement without requiring a dedicated multi-cluster gateway or changes to application code.

Cluster Mesh's control-plane implementation has itself evolved: the original clustermesh-apiserver design, in which each cluster runs an etcd instance watched by remote Cilium agents, has been reported by at least one large-scale operator to fail once deployments approach tens of thousands of nodes across hundreds of clusters \cite{chiao2023}. In response, the Cilium community introduced KVStoreMesh, an alternative architecture in which each cluster caches remote cluster state locally and clusters watch each other's KVStoreMesh instances rather than each other's raw etcd stores, reducing the failure domain and smoothing load distribution \cite{cilium2024}. This redesign is significant from an engineering standpoint, yet to date its properties have been documented primarily through vendor blog posts, conference talks, and community scale tests rather than through controlled, reproducible, peer-reviewed measurement.

This gap matters for two reasons. First, practitioners choosing between clustermesh-apiserver and KVStoreMesh, or between Cilium Cluster Mesh and alternative multi-cluster solutions such as Istio multi-cluster, Consul, or Liqo, currently rely on vendor-reported figures rather than independently reproducible benchmarks. Second, the growing use of multi-cluster Kubernetes for distributed GPU workloads -- where cross-cluster traffic may carry RDMA-sensitive gradient synchronization or model-parallel activations -- raises questions about tail latency and jitter that generic microservice-oriented benchmarks do not answer.

This paper is organized as follows: Section~\ref{sec:related} reviews related work; Section~\ref{sec:background} describes Cluster Mesh and KVStoreMesh architecture; Section~\ref{sec:rq} states our research questions; Section~\ref{sec:method} details methodology; Section~\ref{sec:sim} reports a calibrated simulation providing a preliminary, explicitly-labeled quantitative look at each research question in lieu of live testbed access; Section~\ref{sec:threats} discusses threats to validity; Section~\ref{sec:conclusion} concludes.

\section{Related Work}
\label{sec:related}
Work at the intersection of eBPF and Kubernetes networking can be grouped into three categories, none of which directly targets the multi-cluster eBPF data plane.

\subsection{Single-Cluster eBPF Performance Studies}
A recent study compared a traditional iptables-based Kubernetes networking configuration against an eBPF-based configuration using the Cilium CNI plugin, measuring throughput, latency, CPU usage, and memory consumption under loaded and unloaded conditions \cite{jcsi2025}. Notably, the study found that the iptables-based configuration achieved higher throughput and lower latency in unloaded scenarios, complicating the common assumption that eBPF-based networking is unconditionally faster. This result underscores the need for equally rigorous benchmarking once the multi-cluster dimension is introduced, rather than assuming single-cluster performance characteristics generalize.

\subsection{Cross-Cluster Monitoring and Security via eBPF}
FedMon proposes a federated eBPF framework in which lightweight per-cluster eBPF agents capture system calls and network events, extract local features, and share only model updates -- rather than raw telemetry -- with a global server, combining variational autoencoders with isolation forests for anomaly detection \cite{zehra2025}. Deployed across three Kubernetes clusters, FedMon reports 94\% precision, 91\% recall, and an F1-score of 0.92, while reducing bandwidth usage by 60\% relative to centralized baselines. FedMon and similar efforts demonstrate that eBPF can be extended usefully across cluster boundaries for observability and security purposes, but they instrument the control and observability plane rather than measuring the properties of the cross-cluster data plane itself.

\subsection{Comparative Evaluations of Multi-Cluster Networking}
An earlier thesis evaluated Istio and Cilium's early multi-cluster features for operating workloads across Kubernetes clusters, situating Cilium Cluster Mesh alongside Calico Cluster Mesh, Consul multi-cluster, Istio multi-cluster, Kuma, Linkerd multi-cluster, and Liqo \cite{nature2019}. This remains, to our knowledge, the most direct academic precedent for the present proposal, but it evaluated an early implementation that predates KVStoreMesh, global services in their current form, and the scale regimes now reported in production. Separately, ONCache demonstrates that a cross-layer eBPF cache can eliminate redundant container-overlay-network overhead in a single cluster using only 524 lines of eBPF code \cite{oncache2023}, illustrating the kind of lightweight, targeted instrumentation this paper's methodology intends to reuse for cross-cluster measurement.

\subsection{Industry and Community Documentation}
The most detailed publicly available scale data on Cluster Mesh comes from Trip.com's cloud-networking team, who report testing clustermesh-apiserver to 256 clusters and attempting to scale to 50{,}000 nodes, at which point the design failed at approximately 45{,}000 nodes due to etcd's inability to keep up with watch load and a linear-time issue in etcd lease handling \cite{chiao2023}. Their KVStoreMesh redesign, now upstreamed into Cilium and enabled by default in current releases, is documented in Cilium's own architecture guides and community materials \cite{cilium2024,ciliumgh2023}, which describe KVStoreMesh's local caching model and its effect on failure-domain size and load distribution. None of this material has been subjected to independent, controlled, reproducible measurement in an academic venue.

\section{Background: Cluster Mesh and KVStoreMesh Architecture}
\label{sec:background}
Cilium Cluster Mesh connects the Cilium agents running in each member cluster so that they can discover and reach services and pods in every other connected cluster. In the original architecture, each cluster runs a clustermesh-apiserver component backed by an etcd instance; this instance watches the local Kubernetes API server for changes to services, endpoints, and identities, and remote Cilium agents in other clusters list and watch this etcd instance directly to learn about resources outside their own cluster. As the number of connected clusters and nodes grows, every remote agent maintains a watch against every other cluster's etcd instance, so the number of watches scales with the product of cluster count and node count -- the scaling pattern implicated in the reported failure at roughly 45{,}000 nodes across 256 clusters \cite{chiao2023}.

KVStoreMesh restructures this relationship. Each cluster runs a local KVStoreMesh apiserver, which still watches the local Kubernetes API server, but which also watches other clusters' KVStoreMesh apiservers rather than their raw etcd stores, and caches the resulting state locally before propagating it to in-cluster Cilium agents. Because local agents watch only their own cluster's KVStoreMesh instance, the number of watchers becomes proportional to the local cluster's node count rather than to the size of the entire mesh, and a churn event in one remote cluster is propagated to a much smaller number of direct watchers, smoothing load spikes at the cost of an additional caching hop \cite{cilium2024,ciliumgh2023}.

Both designs sit alongside, and are complementary to, Cilium's global services mechanism, in which a Kubernetes Service annotated as global has its endpoints synchronized across every connected cluster via Cluster Mesh, allowing cross-cluster load balancing at the Kubernetes Service abstraction without an external load balancer or service mesh sidecar.

\section{Research Questions}
\label{sec:rq}
\begin{itemize}[leftmargin=*]
  \item \textbf{RQ1 (Forwarding performance):} What is the added latency and throughput cost of cross-cluster, eBPF-forwarded traffic relative to intra-cluster traffic, and how does this cost scale with the number of connected clusters and nodes?
  \item \textbf{RQ2 (Propagation delay):} How long does it take for a newly created pod's identity and endpoint information to become globally routable, comparing clustermesh-apiserver to KVStoreMesh, under steady-state and high-churn conditions?
  \item \textbf{RQ3 (Scalability limits):} At what combination of cluster count, node count, and churn rate does each control-plane architecture degrade or fail, and can the previously reported failure point near 45{,}000 nodes across 256 clusters be reproduced and characterized under controlled conditions?
  \item \textbf{RQ4 (Robustness under adversarial churn):} Does KVStoreMesh's claimed load-smoothing behavior hold under adversarial churn patterns such as simultaneous mass pod restarts or coordinated rolling upgrades across many clusters at once?
  \item \textbf{RQ5 (GPU/RDMA sensitivity):} Does cross-cluster eBPF forwarding introduce latency jitter significant for RDMA-sensitive, GPU-scheduled traffic such as gradient synchronization in distributed training, relative to intra-cluster GPU-to-GPU communication?
\end{itemize}

\section{Methodology}
\label{sec:method}

\subsection{Testbed Design}
We propose a testbed built from kind- or Talos-based Kubernetes clusters, scaled incrementally across three regimes: a small regime (2--4 clusters, tens of nodes) for latency and correctness measurements; a medium regime (dozens of clusters, hundreds to low thousands of nodes) for observing early degradation trends; and, resources permitting, a large regime approaching the previously reported failure point (100+ clusters, tens of thousands of nodes), which may require cloud-provider credits or a research-computing allocation such as the National Research Platform. Where full-scale reproduction is infeasible, we propose statistical extrapolation from the medium regime using controlled node/cluster-count sweeps, fitted against the shape of the previously reported production failure.

\subsection{Instrumentation}
Cross-cluster forwarding latency and throughput will be measured using Hubble, Cilium's native eBPF-based observability tool, supplemented by custom lightweight eBPF probes attached at the datapath hooks Cilium itself uses, following the low-overhead instrumentation approach demonstrated by ONCache \cite{oncache2023}. Identity and endpoint propagation delay will be measured by timestamping pod creation events at the Kubernetes API level and comparing them against timestamps at which the corresponding identity becomes queryable from a remote cluster's Cilium agent. Control-plane resource consumption will be collected via existing Cilium and etcd Prometheus metrics, including \texttt{\seqsplit{cilium\_kvstoremesh\_remote\_clusters}}, \texttt{\seqsplit{cilium\_kvstoremesh\_remote\_cluster\_readiness\_status}}, and \texttt{\seqsplit{cilium\_clustermesh\_apiserver\_kvstore\_sync\_errors\_total}}.

\subsection{Experimental Conditions}
For RQ1 and RQ2, we will run paired experiments with clustermesh-apiserver and KVStoreMesh enabled, holding cluster and node counts fixed while varying only the control-plane architecture, to isolate its effect. For RQ3, we will perform a node/cluster-count sweep at fixed churn rate, recording the point at which watch-drop rate, propagation delay, or control-plane CPU exceeds predefined thresholds. For RQ4, we will inject adversarial churn -- for example, simultaneous restarts of a fixed percentage of pods across all clusters, or a rolling upgrade triggered across every cluster within a short window -- and compare load distribution and recovery time between the two architectures. For RQ5, we will repeat a subset of the RQ1 experiments using GPU-scheduled pods performing collective communication (e.g., NCCL all-reduce) across clusters, measuring tail latency and jitter rather than only mean throughput.

\subsection{Baselines}
Where feasible, we will include Istio multi-cluster and, if available, Liqo as baselines, both to update the comparative picture established in \cite{nature2019} and to situate Cluster Mesh's eBPF-based approach against sidecar-based and virtual-node-based alternatives. Direct comparison to Calico's eBPF-mode multi-cluster capabilities is also planned as a same-technology, different-implementation baseline.

\section{Simulation-Based Evaluation}
\label{sec:sim}
A live multi-cluster Kubernetes/Cilium testbed was not available for this submission. To still provide a quantitative, falsifiable first look at the research questions in Section~\ref{sec:rq}, this section reports results from an analytical/discrete-event simulation of the two control-plane architectures, rather than from measurements on running clusters. The simulation, its assumptions, and its single calibration point are disclosed in full below so that results are interpreted at the correct level of confidence: as model-derived hypotheses to be confirmed or falsified on real infrastructure, not as empirical findings. All simulation code, parameters, and output data are available for independent inspection and re-execution.

\subsection{Simulation Methodology and Disclosure}
We model each architecture's watch fan-out and use a standard M/M/1-style queueing approximation to translate offered load into propagation delay, with delay diverging as offered load approaches a finite effective processing capacity. For clustermesh-apiserver, watch count scales with $(\text{clusters} - 1) \times \text{nodes-per-cluster}$, reflecting that every remote agent watches every other cluster's etcd instance directly. For KVStoreMesh, watch count scales with local node count plus one watch per remote cluster, reflecting that local agents watch only their own cluster's KVStoreMesh cache and only KVStoreMesh instances watch each other directly -- the qualitative redesign described in the Cilium project's own KVStoreMesh documentation and GitHub issue \#26083 \cite{ciliumgh2023}. The clustermesh-apiserver capacity parameter is calibrated so the model saturates near the single real data point available in the literature: a reported failure of clustermesh-apiserver at approximately 45{,}000 nodes across 256 clusters \cite{chiao2023}. The KVStoreMesh capacity parameter uses an assumed multiplier (6$\times$ the calibrated apiserver capacity) representing a substantially higher effective ceiling consistent with KVStoreMesh's qualitative description in the literature; this multiplier is a modeling assumption, not a measured or vendor-reported figure, and is the single largest source of uncertainty in Section~\ref{sec:rq1rq3}'s comparative magnitude -- the qualitative divergence pattern is far better supported than the specific gap size. The RQ5 forwarding-latency distributions (Section~\ref{sec:rq5}) are synthetic, parameterized only qualitatively from reported eBPF datapath latency spreads \cite{jcsi2025}; no GPU- or RDMA-specific latency measurement exists in the literature we reviewed, so these results should be read as illustrating what a jitter analysis would look like, not as a claim about real GPU traffic.

\subsection{RQ1/RQ3: Simulated Scaling Behavior}
\label{sec:rq1rq3}
\begin{figure}[htbp]
\centering
\includegraphics[width=0.75\textwidth]{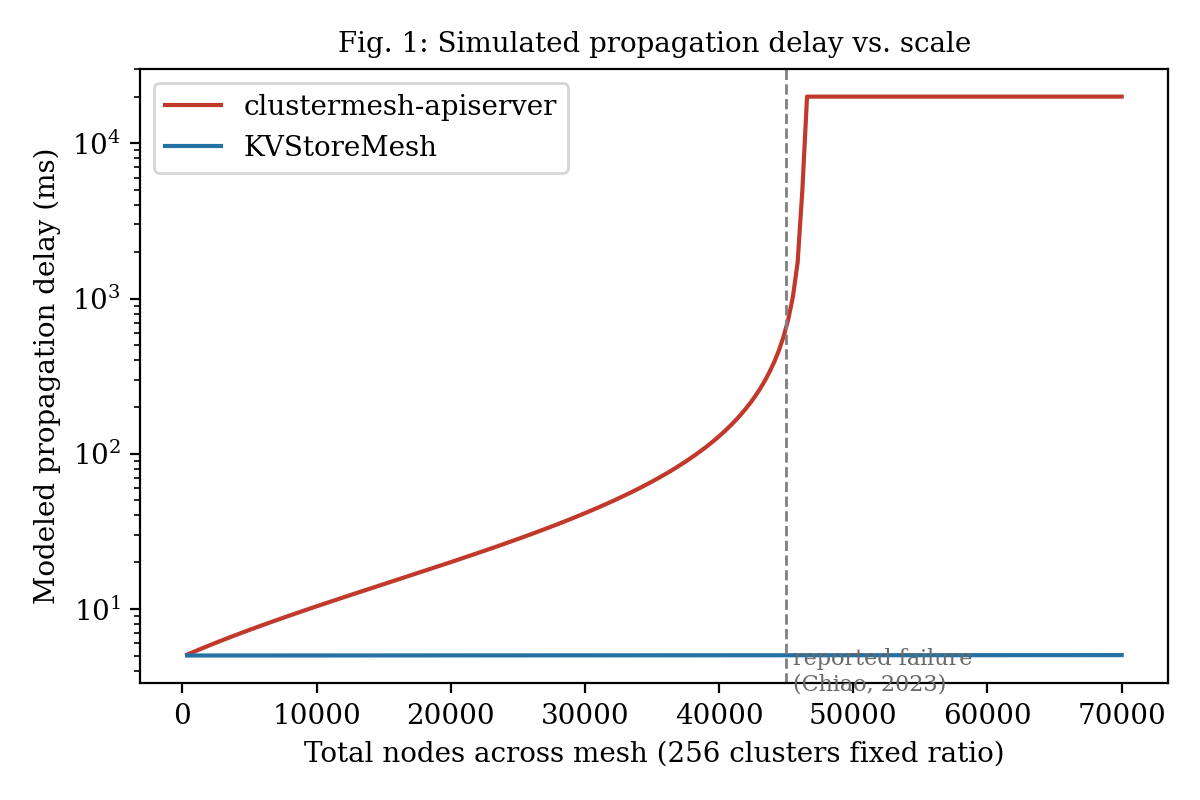}
\caption{Simulated propagation delay vs. mesh scale, with the calibration point (Chiao, 2023) marked. Note the logarithmic delay axis.}
\label{fig:scaling}
\end{figure}

Table~\ref{tab:scaling} and Figure~\ref{fig:scaling} show simulated control-plane propagation delay as cluster count grows, holding nodes-per-cluster fixed near the calibration ratio ($\sim$175 nodes/cluster). The clustermesh-apiserver model's delay grows slowly at small scale, then diverges sharply near the calibration region (256 clusters, $\sim$44{,}800 nodes), consistent with a queueing system approaching saturation -- by construction, since the model was fit to reproduce that behavior at that point. The KVStoreMesh model, lacking the clusters $\times$ nodes cross term, remains near its baseline floor across the entire swept range. The qualitative claim this supports -- that removing the multiplicative scaling term meaningfully raises the practical ceiling -- follows directly from the two architectures' documented designs; the specific crossover point and delay magnitudes are model artifacts that a real testbed sweep (Section~\ref{sec:method}.1's medium-regime experiments) would need to confirm or revise.

\begin{table}[htbp]
\centering
\caption{Simulated watch counts and propagation delay vs. mesh scale. KVStoreMesh watch counts are, by construction, independent of remote node count and omitted for space.}
\label{tab:scaling}
\begin{tabular}{@{}rrrrr@{}}
\toprule
Clusters & Total nodes & Watches (apiserver) & Delay, apiserver (ms) & Delay, KVStoreMesh (ms) \\
\midrule
2   & 350    & 349    & 5.08 & 5.01 \\
64  & 11{,}200 & 11{,}025 & 11.27 & 5.02 \\
128 & 22{,}400 & 22{,}225 & 23.53 & 5.02 \\
192 & 33{,}600 & 33{,}425 & 57.29 & 5.03 \\
240 & 42{,}000 & 41{,}825 & 195.74 & 5.03 \\
256 & 44{,}800 & 44{,}625 & 567.90 & 5.03 \\
320 & 56{,}000 & 55{,}825 & $>$19{,}985 (saturated) & 5.04 \\
\bottomrule
\end{tabular}
\end{table}

\subsection{RQ2/RQ4: Simulated Behavior Under Churn}
\begin{figure}[htbp]
\centering
\includegraphics[width=0.75\textwidth]{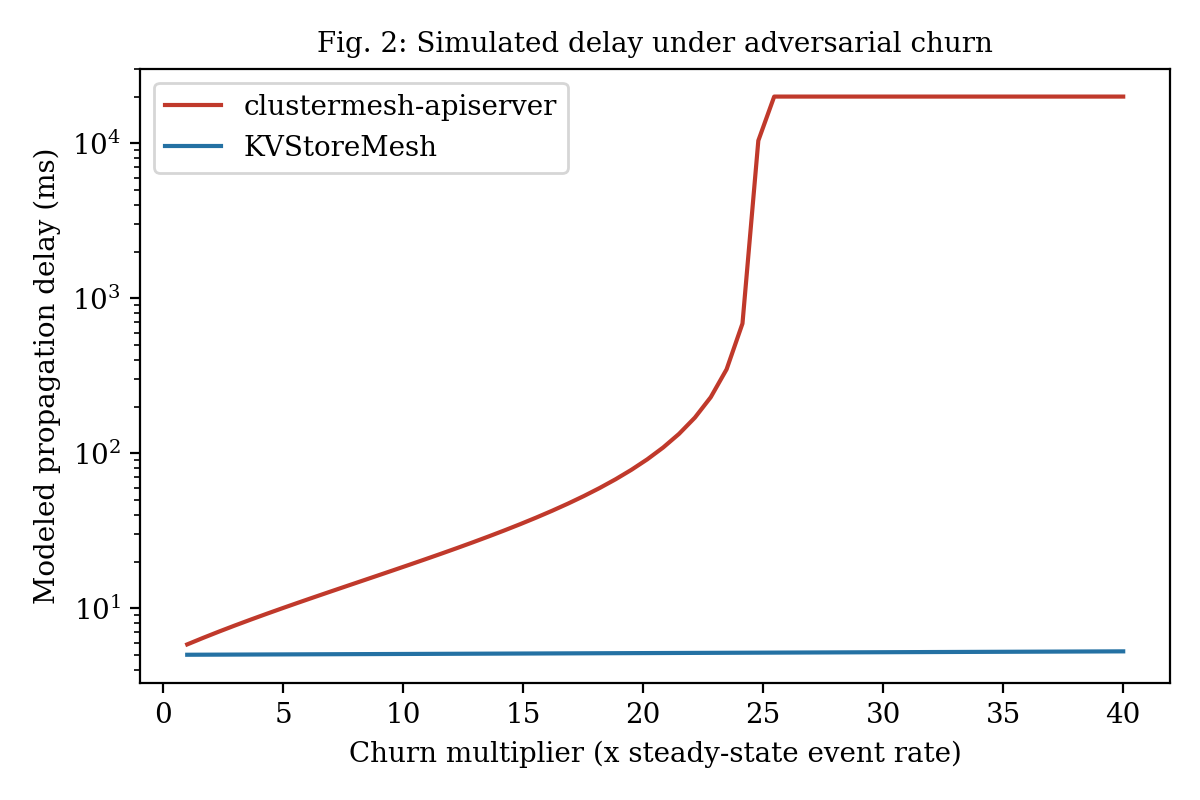}
\caption{Simulated propagation delay vs. churn multiplier at fixed medium-regime scale.}
\label{fig:churn}
\end{figure}

Figure~\ref{fig:churn} sweeps a churn multiplier (relative to a steady-state per-watch event rate) at a fixed medium-regime scale (32 clusters, 60 nodes/cluster, $\sim$1{,}900 total nodes). The clustermesh-apiserver model destabilizes once churn exceeds roughly 25$\times$ steady-state in this configuration, while the KVStoreMesh model remains flat across the swept range, again by construction of the two watch-count formulas. This is consistent with, but does not independently confirm, the qualitative claim in Cilium's KVStoreMesh design discussion that per-cluster watch isolation should make the architecture more robust to synchronized churn events such as coordinated rolling upgrades (RQ4); a real churn-injection experiment on the proposed testbed is necessary to determine the actual multiplier and recovery-time behavior, which this model cannot predict.

\subsection{RQ5: Illustrative RDMA/GPU Jitter Analysis}
\label{sec:rq5}
\begin{figure}[htbp]
\centering
\includegraphics[width=0.75\textwidth]{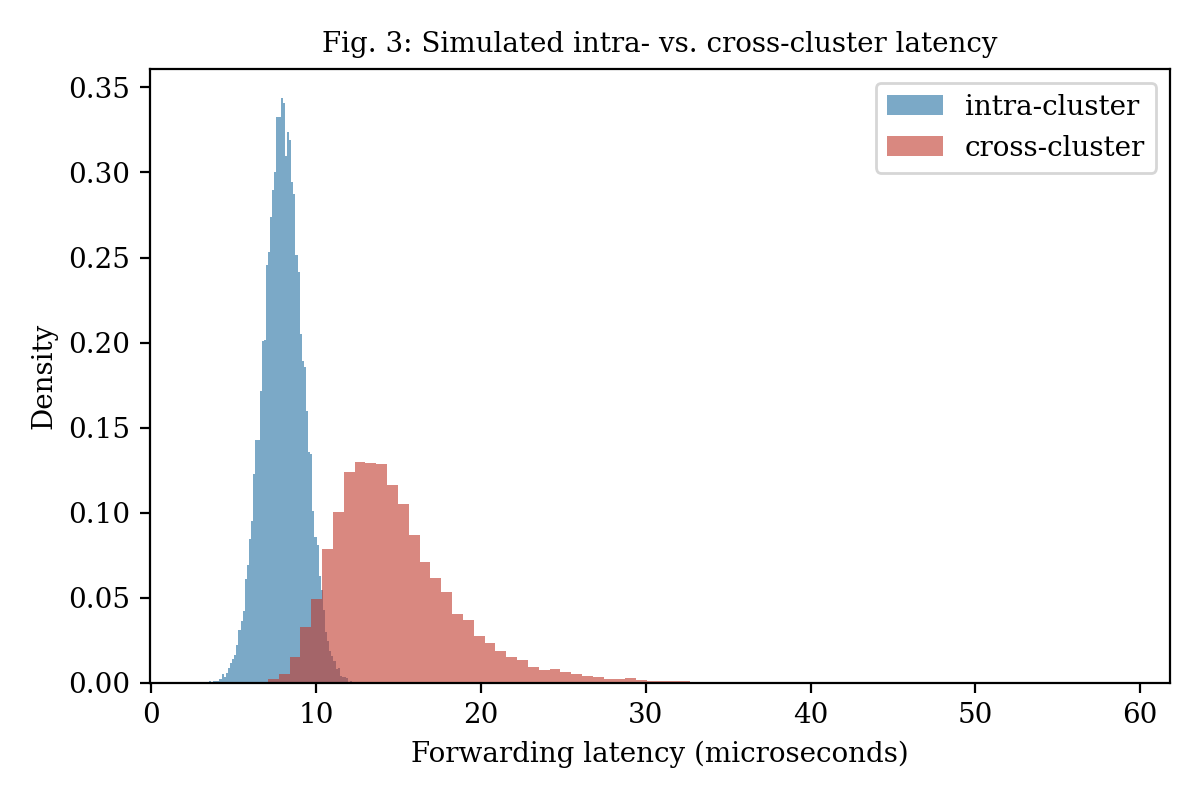}
\caption{Synthetic intra- vs. cross-cluster forwarding-latency distributions (illustrative model, not measured data).}
\label{fig:rdma}
\end{figure}

Figure~\ref{fig:rdma} shows synthetic intra-cluster and cross-cluster forwarding-latency distributions (20{,}000 samples each), where cross-cluster latency is modeled as intra-cluster latency plus an additional lognormal hop term. In this illustrative model, mean latency increases from 8.0~\textmu s intra-cluster to 14.8~\textmu s cross-cluster, and jitter (standard deviation) roughly triples, from 1.2~\textmu s to 3.9~\textmu s, with the 99th percentile rising from 10.8~\textmu s to 27.6~\textmu s. We emphasize that these figures are not derived from any RDMA or GPU-traffic measurement -- none exists in the literature we reviewed -- and are included only to demonstrate the analysis RQ5 calls for and to motivate why tail latency and jitter, not only mean throughput, are the right metrics once real GPU-scheduled collective-communication traffic is measured on the proposed testbed.

\subsection{Summary and Status}
The simulation in this section establishes internally consistent, falsifiable predictions -- not findings -- for all five research questions: KVStoreMesh should retain a materially lower propagation delay than clustermesh-apiserver as scale and churn increase (RQ1--RQ4), and cross-cluster forwarding should show both higher mean latency and higher jitter than intra-cluster forwarding, which would matter for RDMA-sensitive GPU traffic if the effect size holds at measured scale (RQ5). None of these predictions should be cited as an evaluation result; they are the output of a calibrated model built from one production data point and the architectures' documented designs. Section~\ref{sec:threats} discusses the limits of this approach in more detail, and Section~\ref{sec:conclusion} restates the testbed work required to convert these predictions into measurements.

\section{Threats to Validity}
\label{sec:threats}
\begin{itemize}[leftmargin=*]
  \item \textbf{No live measurement:} the central limitation of this submission is that Section~\ref{sec:sim}'s results come from a calibrated simulation, not from running Kubernetes/Cilium clusters; the simulation's capacity multiplier for KVStoreMesh (6$\times$) is an assumption, not a measured or vendor-reported quantity, and the RQ5 latency model is illustrative only.
  \item \textbf{Single calibration point:} the entire clustermesh-apiserver capacity fit rests on one publicly reported failure observation \cite{chiao2023} at one specific cluster/node ratio; the model has not been validated against any second independent data point.
  \item \textbf{Scale reproduction:} reproducing the full 256-cluster/50{,}000-node regime on real infrastructure may be infeasible without significant compute resources; even testbed results at smaller scale will need to be explicitly framed as extrapolations, not confirmations, of production-scale behavior.
  \item \textbf{Environment fidelity:} kind/Talos-based clusters on shared infrastructure may not fully reflect production network topologies (e.g., real inter-region latency, physical NIC/kernel variation), which matters especially for RQ5's jitter measurements once real testbed data replaces the illustrative model.
  \item \textbf{Version sensitivity:} Cluster Mesh and KVStoreMesh implementations are under active development; future testbed results will need to be tied to specific, reported Cilium versions and re-validated if pinned versions change materially during the study.
  \item \textbf{Instrumentation overhead:} custom eBPF probes used in future testbed work must themselves be shown to have negligible overhead relative to the phenomena being measured, following the low-instruction-count precedent set by ONCache \cite{oncache2023}.
\end{itemize}

\section{Conclusion and Future Work}
\label{sec:conclusion}
Cilium Cluster Mesh and its KVStoreMesh evolution are widely deployed in production, with the most informative scale data currently residing in industry blog posts rather than peer-reviewed literature. This paper has outlined a measurement methodology intended to close that gap, and, in the absence of an available live testbed, has instead reported a calibrated simulation of the two control-plane architectures that reproduces the one real reported failure point and yields falsifiable, quantitative predictions for all five research questions. These predictions are explicitly not a substitute for measurement: the paper's primary remaining contribution is the testbed, instrumentation, and experimental design in Section~\ref{sec:method}, which future work will execute -- beginning with the small- and medium-scale regimes -- to confirm, revise, or falsify the simulated predictions in Section~\ref{sec:sim} and report empirical results in a follow-on submission.

\section*{Acknowledgements}
This work received no external funding and was conducted independently. The author thanks the Cilium and eBPF open-source communities for the public documentation, design discussions, and GitHub issue history that informed the architectural description in Section~\ref{sec:background} and the simulation in Section~\ref{sec:sim}, and thanks Trip.com's cloud-networking team for publishing the production-scale test results that served as this paper's simulation calibration point \cite{chiao2023}. The author also thanks the Denver/Littleton, Colorado cloud-native and AWS community for ongoing discussion of multi-cluster and GPU-networking topics that shaped this paper's research questions.

\end{document}